\documentclass[nofootinbib,twocolumn,aps,pre,superscriptaddress,citeautoscript,floatfix]{revtex4-2}
\usepackage{graphics}
\usepackage{graphicx}
\usepackage{color}
\usepackage{amsmath}
\usepackage{amssymb}
\usepackage{float}
\newcommand\be{\begin{equation}}
\newcommand\ee{\end{equation}}
\newcommand\bea{\begin{eqnarray}}
\newcommand\eea{\end{eqnarray} }

\begin{document}

\title{Phase Separation of Polyelectrolytes: \\ The Effect of Charge Regulation}
\author{Bin Zheng}
\affiliation{Raymond and Beverly Sackler School of Physics and Astronomy,
Tel Aviv University, Ramat Aviv 69978, Tel Aviv, Israel}
\author{Yael Avni}
\affiliation{Raymond and Beverly Sackler School of Physics and Astronomy,
Tel Aviv University, Ramat Aviv 69978, Tel Aviv, Israel}
\author{David Andelman$^*$}
\footnotetext{andelman@post.tau.ac.il}
\affiliation{Raymond and Beverly Sackler School of Physics and Astronomy,
Tel Aviv University, Ramat Aviv 69978, Tel Aviv, Israel}
\author{Rudolf Podgornik$^*$}
\footnotetext{podgornikrudolf@ucas.ac.cn \\Also affiliated to: Department of Physics, Faculty of Mathematics and Physics, University of Ljubljana, 1000 Ljubljana, Slovenia}
\affiliation{School of Physical Sciences and Kavli Institute for Theoretical Sciences, University of Chinese Academy of Sciences, Beijing 100049, China}
\affiliation{Wenzhou Institute of the University of Chinese Academy of Sciences, Wenzhou, Zhejiang 325000, China}
\affiliation{CAS Key Laboratory of Soft Matter Physics, Institute of Physics, Chinese Academy of Sciences, Beijing 100190, China}


\begin{abstract}
Complex coacervation, known as the liquid-liquid phase separation of solutions with oppositely charged polyelectrolytes, has attracted substantial interest in recent years. We study the effect of the charge regulation (CR) mechanism on the complex coacervation by including short-range interactions between the charged sites on the polymer chains as well as an association-dissociation energy parameter in the CR mechanism.  We investigate the phase diagrams of two CR models: (i)~the hopping CR model (HCR) and (ii)~the asymmetric CR model (ACR).  It is shown that during the phase separation that the polymers in the condensed phase are more charged than those in the dilute phase, in accordance with Le Chatelier's principle. In addition, {\it secondary CR} effects also influence the change in the volume fraction of the two phases. The latter can cause the charge difference between the two phases to change nonmonotonically as a function of the CR parameters.
\end{abstract}

\maketitle


\section{Introduction}
Solutions of oppositely charged polyelectrolytes under certain conditions can undergo a liquid-liquid phase separation, resulting in a condensed phase coexisting with a dilute one. This phenomenon was described almost a century ago~\cite{Bungenberg} and is known as {\it complex coacervation}. It has instigated broad interest in many areas of soft matter science, including polymers, colloids and protein physics~\cite{{Charles2020,Rumyantsev2021}}. More recently, complex coacervation was invoked as a potential mechanism underpinning the formation of membrane-less intracellular compartments in biological systems, playing a key role in controlling biochemical processes within the cell~\cite{Brangwynne2011, Brangwynne2015}.

Complex coacervation was first modeled and analyzed by Voorn and Overbeek~\cite{voorn1956}. In the Voorn-Overbeek (VO) model, the phase separation results from a competition, described within  the Debye-H\"uckel (DH) theory~\cite{Debye1923}, between the entropy of mixing and the electrostatic fluctuation attraction between charged monomers. The VO model largely captures the phenomenology of coacervate phase behavior, although the model neglects the connectivity of the polymer chains and its validity conditions are limited by DH theory to low salt concentrations. Later on, different generalizations of the model as well as improvements in its methodology and computational aspects have been proposed~\cite{Wang2018,Charles2020}. These later works capture potentially relevant physics that the VO model might have missed  and are based on  various aspects of polymer field-theory, scaling theory, and counterion release models.

Unlike the assumed constant charge assigned to the polymer chains in the original VO model, the charge association/dissociation processes of chargeable polymer groups lead to the {\it charge regulation} (CR) mechanism~\cite{yael2018}. The principal effect is that the variation in the polymer charge is a function of the polymer concentration, added salt concentration, and/or solution pH. The CR mechanism was first invoked to describe the acid/base properties of polyelectrolytes as well as the dissociation equilibria of proteins. (For historical references, see ref.~\cite{Yael}).

In the 1970s Ninham and Parsegian~\cite{ninham1971}  formulated the CR mechanism within the Poisson-Boltzmann (PB) theory, specifically in the context of membrane interactions in multilamellar assemblies. The importance of the CR mechanism in explaining properties of protein solutions is well recognized~\cite{Lund} and  is invoked regularly to address the polymer-charge variation under various solution conditions~\cite{Yael}. Surprisingly, the CR mechanism has not been regularly employed in the modeling of coacervation phenomenology, and pertinent analyses are rather scarce~\cite{Charles2020}.

Among the works addressing  CR within the context of the liquid-liquid phase-separation of polyelectrolytes, we specifically mention two that are of great relevance to the present work. Muthukumar {\it et al.}~\cite{Muthukumar2010} studied CR in salt-free polyelectrolyte solutions within a generalized VO free-energy model by considering a single type of negatively charged polymer chain and its positively charged counterions. It was shown that the polymer charge is self-regulated during the phase separation, contrary to the original VO assumption of fixed polymer charge.  Furthermore, this implies that the coexisting phases possess different degrees of ionization, with the more condensed polymer phase having a smaller amount of charge on its polymer chains.

Salehi and Larson~\cite{Larson2016} studied a more complex system and considered three different types of short-range electrostatic effects. In their study, they included the association/dissociation of the CR counterions and ion pairing of charges on oppositely charged polyelectrolyte and based their model on an extended version of the VO free energy. The weak polyelectrolyte phase separation was shown to be quite sensitive to the solution pH. It revealed that the complex coacervation of charged polymers can be simply explained by the competition between counterion condensation and cross-chain ion pairing.

Complexation in solutions of oppositely charged polyelectrolytes including polyion cross-linking due to formation of thermoreversible ionic pairs was pursued also in the work of Kudlay et al.\cite{Kudlay2004}. Here the degree of ion association was obtained self-consistently from a mass action law, similar to charge regulation, and the complexation was shown to be susceptible to long-range electrostatics, ion association, as well as van der Waals attraction.

In the works mentioned above~~\cite{Muthukumar2010,Larson2016}, the effects of CR on the complex coacervation were studied for specific models, but some of the more interesting possible behaviors of CR-induced phase-separation were not explored. Moreover, the CR mechanism was formulated within a Langmuir adsorption isotherm model that is based on an association-dissociation energy cost and a lattice-gas adsorption entropy. This implies that the adsorption process is neither subject to short-range interactions between the occupied sites, nor exhibits any cooperativity. Including short-range interactions leads, in general, to an adsorption isotherm of the Frumkin-Fowler-Guggenheim variety~\cite{Koopal} that has different properties, which can fundamentally change the adsorption phenomena~\cite{Majee,Avni2020}.
It is therefore of importance to explore the more complicated adsorption isotherms, specifically as they relate to  phase separations. 

In the present study, we focus on the effect of CR on the phase separation of polyelectrolyte solutions by considering the role of two separate interaction mechanisms on the adsorption process. (i)~the free-energy change in the adsorption/desorption process
of a single adsorption site, quantified by a parameter $\alpha$, and (ii)~the free-energy change due to the short-range interaction between two charged neighboring adsorption/desorption sites, as quantified by another parameter $\eta$.

We specifically address two variations of the standard CR model. (i)~The first is called the {\it hopping CR model} (HCR), where we start with two polymer types without any dissociated charge groups.
Because of the adsorption/desorption of ionic groups on the chains, the charges released from one polymer type are immediately captured by the other polymer, which favors being oppositely charged. (ii)~The second model is the {\it asymmetric CR model} (ACR), in which one polymer type has a constant negative charge ({\it i.e.}, the charges are completely dissociated), while the counterions can be either free in solution or adsorbed onto the other type of polymer chains.

We investigate the effects of CR on the polyelectrolyte phase diagram and analyze the variation of the polymer charge for the HCR/ACR models. Unlike the observation by Muthukumar {\it et al}~\cite{Muthukumar2010}, we find that the polymers
in the condensed phase are more highly charged than in the coexisting dilute phase. This is due to the different ionization processes of the two polymers and is also consistent with Le Chatelier's principle. We then show that the two CR parameters can affect the polymer charge of the two phases in a nonmonotonic way. Upon variation of a single CR parameter, the charge asymmetry between the phases first increases and then decreases. This is explained in terms of a {\it secondary CR effect}, where the polymer charge is regulated directly from the adsorption/desorption chemical reactions as it is in a single phase but is also regulated indirectly from the change in volume fraction of the two phases, which is also caused by CR.

The outline of the article is as follows. In section II, we introduce the general model and its pertinent free-energy, and focus on two variants of the CR model. In section III, we discuss the effects of the CR on the complex coacervation and of the ionization states of the two phases. Finally, section IV includes some suggestions for future experiments and our conclusions.

\section{Model}
The model system under consideration contains two types of polymer chains, each having $N$ monomers per chain, counterions and solvent (water). The two types of polymers are polycations and polyanions, denoted by $p$ and $n$, respectively. The monomers, counterions and water molecules are assumed to have the same molecular volume, $v$. A fraction of monomers, $0\le \gamma\le 1$, contains ionizable groups that can undergo a chemical reaction and become charged. We denote these groups (hereafter referred as ``sites") on the $p$ polymers by A  and on the $n$ polymers by BH. The A sites can become positively charged by association,
\be
\label{eAH}
{\rm A + H^{+} ~~\rightleftarrows ~~AH^{+}.}
\ee
while the B sites can become negatively charged by dissociation,
\be
\label{eBH}
{\rm BH~~ \rightleftarrows ~~B^{-}+H^{+}.}
\ee
The total charge on each polycation chain is $z_{+}$ and for the polyanions it is $z_{-}$. Note that $z_{+}$ and $z_{-}$ are not fixed but are annealed (adjustable) parameters, and their maximal value is $z_0=\gamma N$.

The model free energy consists of three separate contributions: a polymer term $f_{\rm P}$, a CR term $f_{\rm CR}$, and an electrostatic term $f_{\rm EC}$
\be
F=\int_V f\,{\rm d}V~=\int_V\left(f_{\rm P}+f_{\rm CR}+f_{\rm EC}\right)\,{\rm d}V~.
\label{e1}
\ee

The general form of the dimensionless free-energy density per site is given by the Flory-Huggins free-energy,
\be
\begin{aligned}
\frac{v}{k_{\rm B}T}\displaystyle f_{\rm P}=& {\frac {\phi_{p} }{N}}\ln \phi_{p} + {\frac {\phi_{n} }{N}}\ln \phi_{n}+  {\phi_w }\ln \phi_w  \\
&+  {\phi_{\rm ci}}\ln \phi_{\rm ci} + \chi_{+} \phi_{p}\phi_w + \chi_{-}\phi_{n} \phi_w .
\end{aligned}
\label{e2}
\ee
The parameters that appear in eq~\ref{e2} are  the thermal energy $k_{\rm B}T$, and the volume fractions of the $p$ and $n$ polymers, $\phi_{p,n}$, related by $\phi_{p,n}=n_{p,n} N/N_{\rm tot}$ to the number of $p$ and $n$ chains, $n_{p,n}$, and to the total number of sites in the system, $N_{\rm tot}$. The other two volume fractions are those of the counterions and water molecules,  $\phi_{\rm ci}$ and $\phi_{w}$. The incompressibility condition related the four volume fractions in the system:
\be
\phi_{p} + \phi_{n} + \phi_w + \phi_{\rm ci}= 1.
\label{e3}
\ee

Generally, the two-body interaction should include all of the species [{\it i.e.}, $\frac{1}{2}\sum_{i,j}\chi_{ij}\phi_{i}\phi_{j}$ ($i=$ polymers, counterions, and solvent)], but here we assume that all of the $\chi_{ij}$ terms, except the interactions between the solvent molecules ($w$) and the two types of polymer segments ($p$ and $n$), are negligible as compared to the electrostatic interaction between the polymer chains.
Hence, we are left only with the two interaction parameters: the polycation-water ($\chi_{+}=\chi_{\rm pw}$) and the polyanion-water ($\chi_{-}=\chi_{\rm nw}$).
With these conditions and definitions, the above free-energy, $f_{\rm P}$, is reduced to
\be
\begin{aligned}
\frac{v}{k_{\rm B}T}\displaystyle f_{\rm P} = &{\frac {\phi_{p}}{N}}\ln \phi_{p} + {\frac {\phi_{n}}{N}}\ln \phi_{n} + \phi_{\rm ci}\ln\phi_{\rm ci} \\
&+ {(1-\phi_{p}-\phi_{n}-\phi_{\rm ci})}\ln(1-\phi_{p}-\phi_{n}-\phi_{\rm ci}) \\
&+(\chi_{+} \phi_{p}+\chi_{-}\phi_{n}) (1-\phi_{p}-\phi_{n}-\phi_{\rm ci}).
\label{e4}
\end{aligned}
\ee

For the electrostatic free energy, $f_{\rm EC}$,
we treat the charges on the polymers as free ions and neglect the chain connectivity, as was done in the extended VO model. The bulk polymer solution is overall neutral and the mean electric field is zero on average everywhere. Consequently, the electrostatic energy also vanishes to the lowest order, and the first correction term, due to Gaussian fluctuations around the zero average potential, is the DH correlation term \cite{Debye1923,Falkenhagen1971,Levin2002}
\be
\label{e9}
\frac{v}{k_{\rm B}T }\displaystyle f_{\rm EC}=-\frac{1}{4\pi}\left(\ln(1+\tilde\kappa)-\tilde\kappa+\frac{1}{2}\tilde\kappa^2\right)
\ee
where
\be
\label{e9n}
\tilde\kappa^2=(\kappa a_w)^2=\lambda \left(\frac{\phi_{p}z_{+}+\phi_{n}z_{-}}{N} +
\phi_{\rm ci}\right)
\ee
and $\lambda=4\pi l_{\rm B}/a_w$, $a_w=v^{1/3}\simeq 3.11{\rm \AA}$ is the cube root of one water molecule's volume $v$, $l_{\rm B}=e^2/(\varepsilon k_{\rm B} T)\simeq 7\,{\rm \AA}$ is the Bjerrum length in water, and thus $\lambda=26.68$ is taken as a dimensionless constant hereafter.

The CR free-energy density per site can be written as~\cite{yael2018}:
\be
\label{e5}
{v} \displaystyle f_{\rm CR} = \frac{\phi_{p}}{N}g_{p}(z_{+}) + \frac{\phi_{n}}{N}g_{n}(z_{-}),
\ee
where $g_{p}$ and $g_{n}$ are the CR free-energies, respectively, of a single polycation and a polyanion~\cite{Yael}. These free energies contain contributions of a single ion adsorption/desorption to/from a single site, the short-range pair interaction between charged neighboring sites and the lattice-gas entropy.

With this in mind,  $g_{p}$ and $g_{n}$ take the forms~\cite{yael2018,Yael}
\bea
\label{e6}
\frac{g_{p}(z_{+})}{k_{\rm B}T} &=& \alpha_{+} z_{+}+\frac{\eta_{+}}{z_0}z_{+}^2
  +\,z_{+}\ln z_{+} \nonumber\\
  &+&(z_0-z_{+})\ln(z_0-z_{+}) \, ,
\eea
and
\bea
\label{e7}
\frac{g_{n}(z_{-})}{k_{\rm B}T} &=& \alpha_{-} z_{-}+\frac{\eta_{-}}{z_0}z_{-}^2
  +\, z_{-}\ln z_{-} \nonumber\\
  &+&(z_0-z_{-})\ln(z_0-z_{-}) \, ,
\eea
where $\alpha_{\pm}$ parameterizes the free-energy change in adsorbing/desorbing an ion to/from a single site and $\eta_{\pm}$ is the change in the free energy due to short-range interactions between neighboring charged adsorption sites. Finally, the last two terms describe the entropy, which accounts for the number of different ways to have $z_{\pm}$ charged sites out of the total number of sites, $z_0=\gamma N$.

Taking into account eqs~\ref{e4}-\ref{e7}, the total free-energy density $f$ per single dissociable site can finally be written as
\be
\begin{aligned}
\frac{v}{k_{\rm B}T}\displaystyle f =&{\frac {\phi_{p}}{N}}\ln \phi_{p} + {\frac {\phi_{n}}{N}}\ln \phi_{n} + \phi_{\rm ci}\ln{\phi_{\rm ci}} \\
&+ {(1-\phi_{p}-\phi_{n}-\phi_{\rm ci})}\ln(1-\phi_{p}-\phi_{n}-\phi_{\rm ci}) \\
&- \frac{1}{4\pi}\Big[\ln(1+\tilde\kappa)-\tilde\kappa+\frac{1}{2}\tilde\kappa^2\Big] \\
&+ (\chi_{+} \phi_{p}+\chi_{-}\phi_{n}) (1-\phi_{p}-\phi_{n}-\phi_{\rm ci})\\
&+ \frac{\phi_{p}}{N}\Big[\alpha_{+} z_{+}+\frac{\eta_{+}}{z_0}z_{+}^2 +z_{+}\ln z_{+}  \Big]\\
&+ \frac{\phi_{p}}{N}\Big[ (z_0-z_{+})\ln(z_0-z_{+})\Big] \\
&+ \frac{\phi_{n}}{N}\Big[\alpha_{-} z_{-}+\frac{\eta_{-}}{z_0}z_{-}^2 + z_{-}\ln z_{-} \Big]\\
&+ \frac{\phi_{n}}{N}\Big[(z_0-z_{-})\ln(z_0-z_{-})\Big].
\end{aligned}
\label{et}
\ee
The $z_{+}$ and $z_{-}$ charges are annealed variables that can be adjusted by the CR process.  At thermodynamical equilibrium, their value is  determined from the minimum condition of the free energy, $\partial f/\partial z_{\pm}=0$. Therefore, $z_{+}$ and $z_{-}$ are functions of the three volume fractions, $\phi_{p}$, $\phi_{n}$, and $\phi_{\rm ci}$, although in most cases their functional dependence cannot be expressed explicitly.

In the following section, two simplified variants of the CR model will be presented separately.

\subsection{Hopping CR Model}

\begin{figure}[h]
\centering
{\includegraphics[width=0.47\textwidth,draft=false]{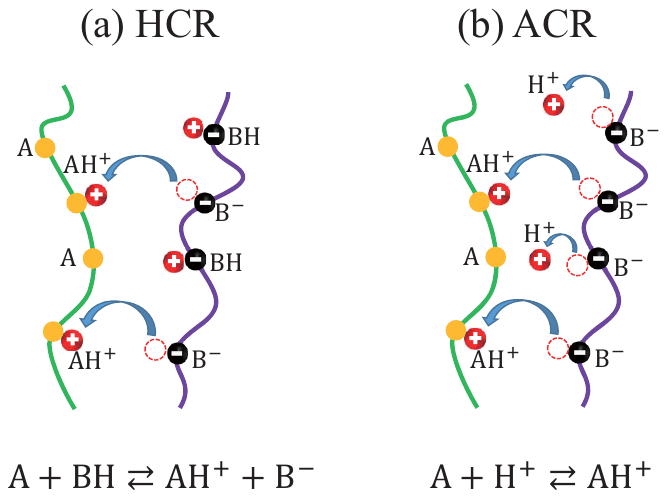}}
\caption{
\textsf{Schematic drawing of the two variants of the CR model. (a)~In the hopping CR model (HCR), the charges dissociated from the B sites on one polymer type are immediately captured by the A sites on the other polymer. (b)~In the asymmetric CR model (ACR), the charges of the B sites dissociate into the solution, and only part of them are captured by the A sites on the other polymer.
}}
\label{fig1}
\end{figure}

We first consider the hopping CR model (HCR) where the charges released from the $n$ sites will be immediately captured by the $p$ sites and vice versa. Hence, one can think of the ions as hopping from one polyion to another. The reactions in eqs~\ref{eAH} and \ref{eBH}
are then reduced to a single reaction,
\be
\label{e12}
\begin{aligned}
{\rm A + BH~~ \rightleftarrows ~~AH^{+}  + B^{-}.}
\end{aligned}
\ee
In this HCR model (see Figure~\ref{fig1}a), there are no free counterions in solution, and the system contains three components: two types of polymer chains and water. Since there is complete symmetry between the $p$ and $n$ polymer types, we can write $\alpha\equiv \alpha_{+} = \alpha_{-}$, $ \chi \equiv \chi_{+} = \chi_{-}$, $\eta\equiv \eta_{+} = \eta_{-}$, $z\equiv z_{+}=z_{-}$, and $\phi= 2\phi_{p} = 2\phi_{n}$. In other words,
$\phi=(n_{n}+n_{p})N/N_{\rm tot}$ is the total polymer volume fraction.

The free-energy density per site, eq~\ref{et}, now has the simplified form,
\be
\label{e13}
\begin{aligned}
\frac{v}{k_{\rm B} T}\displaystyle f=
&-\frac{1}{4\pi}\ln\left[1+\left(\frac{\lambda\phi z}{N}\right)^{1/2}\right]\\
&+\frac{1}{4\pi}\left(\frac{\lambda \phi z}{N}\right)^{1/2}-
~\frac{1}{8\pi}\frac{\lambda \phi z}{N}\\
&+\frac{\phi}{N}\ln{\frac{\phi}{2}} + (1-\phi)\ln{(1-\phi)} \\
&+\frac{\phi}{N}z\ln{z}+\frac{\phi}{N}(z_0-z)\ln{(z_0-z)}\\
&+\chi\phi(1-\phi)+\alpha \frac{\phi}{N}z+\eta\frac{\phi}{Nz_0}z^2.
\end{aligned}
\ee
Minimizing $f$ with respect to $z$, $\partial f/\partial z=0$, leads to the relation
\be
\label{e14}
\begin{aligned}
\alpha+\eta\frac{2z}{z_0}-\frac{\lambda/(8\pi)}
{1+\sqrt{N/( \lambda \phi z)}}-\ln(z_0-z)+\ln z=0,
\end{aligned}
\ee
which is an implicit relation for $z=z(\phi)$.

The phase separation for the polymer/polymer/water system between two coexisting phases is investigated by the usual common tangent construction,
\be
\label{e10}
\frac{\partial f\left(\phi_{1}\right)}{\partial\phi_{1}}=\frac{\partial f\left(\phi_{2}\right)}{\partial\phi_{2}}=
\frac{f\left(\phi_{2}\right)-f\left(\phi_{1}\right)}{\phi_{2}-\phi_{1}},
\ee
where $\phi_1$ and $\phi_2$ are the two coexisting volume fractions on the
binodal (coexisting curve).
In addition, the critical point is determined by $\partial^2 f/\partial^2 \phi =\partial^3 f/\partial^3 \phi =0$.

\subsection{Asymmetric CR Model}

Next, we consider the asymmetric CR model (ACR), where the B sites on the polyanion are fully dissociated, such that there is only one relevant chemical reaction for the A sites (see Figure~\ref{fig1}b),
\be
\label{eAH1}
\begin{aligned}
{\rm A + H^{+}~~ \rightleftarrows ~~ AH^{+}.}
\end{aligned}
\ee
This corresponds to the limit $\alpha_{-}\to -\infty$, where the $n$ polymers have a constant charge on their chain, $z_{-}=z_0$, and $z_{+}$ is an annealed thermodynamic variable that can be adjusted on the $p$ polymers. From electroneutrality, the following condition must hold
\be
\label{e18}
\phi_{\rm ci}=\frac{\phi_{n}z_0}{N}-\frac{\phi_{p}z_{+}}{N}.
\ee
where $\phi_{\rm ci}$ is the counterion volume fraction.

As was done in the previous HCR case, we study here for the ACR case only the symmetric situation, for which $\phi_{p}=\phi_{n}=\phi/2$. The resulting total free-energy density per site eq~\ref{et} is expressed as
\be
\label{e19}
\begin{aligned}
\frac{v}{ k_{\rm B}T }\displaystyle f=
&-\frac{1}{4\pi}\left[\ln\left(1+\tilde\kappa\right)-\tilde\kappa+\frac{1}{2}\tilde\kappa^2\right] \\
&+\frac{\phi}{N}\ln\frac{\phi}{2}+\phi_{\rm ci}\ln\phi _{\rm ci}+\phi_w\ln\phi_w+\chi\phi\phi_w  \\
&+\alpha \frac{\phi  z_{+}}{2N} + \eta \frac{\phi z_{+}^2}{2Nz_0} + \frac{\phi z_{+}}{2 N}\ln z_{+}\\
&+\frac{\phi}{2 N}(z_0-z_{+})\ln(z_0-z_{+}),
\end{aligned}
\ee
where $\tilde{\kappa}$ in eq~\ref{e9n} is adapted for the ACR model
\be
\label{e19n}
\tilde\kappa^2=\lambda\left(\frac{\phi(z_{+} + z_0 )}{2N}+ \phi_{\rm ci}\right)=\lambda\frac{\phi z_0}{N}.
\ee
Substituting the electroneutrality condition (eq~\ref{e18}) and eq~\ref{e19n} into eq~\ref{e19} yields,
\be
\label{e20}
\begin{aligned}
\frac{v}{k_{\rm B} T }\displaystyle f=
&-\frac{1}{4\pi}\left[\ln\left(1+\left(\lambda\frac{\phi z_0}{N}\right)^{1/2}\right) - \left(\lambda\frac{\phi z_0}{N}\right)^{1/2} + \lambda\frac{\phi z_0}{2N}\right] \\
&+\frac{\phi}{N}\ln\frac{\phi}{2} + \frac{\phi}{2N}(z_0-z_{+})
\ln\left[\frac{\phi}{2N}(z_0-z_{+})\right]      \\
&+\left(1-\phi-\frac{\phi z_0}{2N}+\frac{\phi z_{+}}{2N}\right)
\ln\left(1-\phi-\frac{\phi z_0}{2N}+\frac{\phi z_{+}}{2 N}\right)    \\
&+\chi \phi \left(1-\phi-\frac{\phi z_0}{2N}+\frac{\phi z_{+}}{2 N}\right)  \\
&+\alpha \frac{\phi z_{+}}{2N}+\eta \frac{\phi z_{+}^2}{2Nz_0}
+\frac{\phi z_{+}}{2 N}\ln z_{+}\\
&+\frac{\phi(z_0-z_{+})}{2 N}\ln(z_0-z_{+}).
\end{aligned}
\ee

Similar to what was done for the HCR model, eq~\ref{e14}, we minimize $f$ with respect to $z_{+}$, $\partial f/\partial z_{+}=0$, and obtain the following implicit relation, $z_{+}=z_{+}(\phi)$,

\be
\label{e21}
\begin{aligned}
\alpha+2\eta\frac{z_{+}}{z_0}+\chi\phi+\ln{\left[\frac{z_{+}(2N(1-\phi)-\phi(z_0-z_{+}))}{ \phi(z_0-z_{+})^2}\right]}=0.
\end{aligned}
\ee
The phase diagram is obtained by the common-tangent construction, eq~\ref{e10}.

\bigskip
\bigskip
In the two models considered above, polyanions are assumed to be the only source of protons in the system. We do not take into account the protons resulting from the self-dissociation of water molecules yielding its canonical pH. However, for nondilute polymer solutions, this contribution is small compared to the contribution of dissociated polyanions. Additionally, as our main focus is to show clearly the effects of the charge-regulation parameters on the phase diagram, the addition of salt is neglected.

\section{Results and Discussion of the Phase Behavior}

\subsection{The $z(\phi)$ Dependence in the Single Phase}

\begin{figure}[h]
\centering
{\includegraphics[width=0.45\textwidth,draft=false]{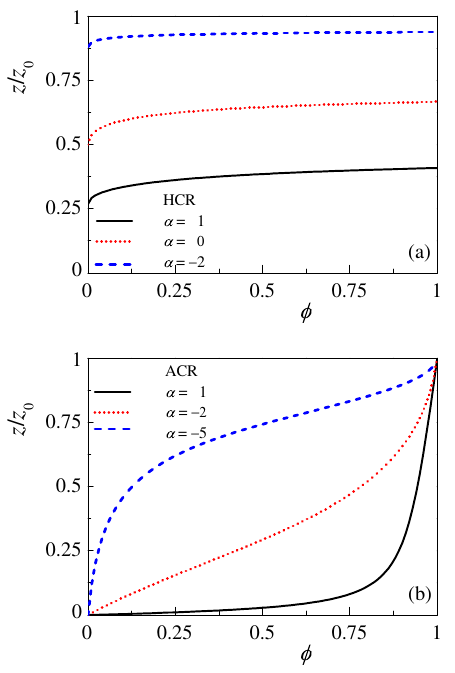}}
\caption{
\textsf{Plots of $z/z_0$ as function of $\phi$  for (a)~the symmetric hopping CR model (HCR), $z\equiv z_{+}=z_{-}$ and $\alpha=1,0$, and $-2$. (b)~Asymmetric CR model (ACR), where $z\equiv z_{+}$, for $\alpha=1, -2,$ and $-5$.  The other parameters are $\eta=0$, $\gamma=0.2$,
$N=200$, $z_0=\gamma N=40$,
$\chi=0.6$, and $\lambda=26.68$. Note that for $\eta=0$ in the HCR model, $z/z_0\to 1/(1+{\rm e}^\alpha)$ when $\phi\to 0$, while for the ACR model, $z/z_0\to 0$ for $\phi\to 0$.}}
\label{fig2}
\end{figure}

We assume that all polymer chains are composed of $N=200$ monomers, and $20\%$ of them ($\gamma = 0.2$) have ionizable groups that can undergo a chemical reaction and become charged. The fraction of charged segments, $z/z_0$, is charge-regulated during the phase separation. Before showing the phase diagrams, we first analyze the dependence of $z$ on $\phi$, as is derived from eqs~\ref{e14} and \ref{e21} for the two models and is shown in Figure~\ref{fig2}. For simplicity, we consider hereafter  $z\equiv z_{+}=z_{-}$ in the symmetric HCR model and  $z\equiv z_{+}$ and $z_{-}=z_0$ in the ACR model.

The $z(\phi)$ dependence indicates that in both HCR and ACR models the polymer charge increases with the polymer concentration. In the HCR model, the coupling between $z$ and $\phi$ is due to the DH correlation term (eq~\ref{e14}), which due to its nonlinearity, favors higher charge at higher polymer concentrations. Because of the hopping of ions between the polymers, the entropy of the free species is not coupled to the polymer charge.

%

The situation is reversed in the ACR model. The DH correlation term does not determine $z$ because the system bears an equal amount of charge when the $\rm{H}^+$ ions are adsorbed onto the polymer, or when they stay in solution. However, because of the electroneutrality (eq~\ref{e18}), the entropy and the short-range interaction between the polymer and the solvent are coupled to $z$ (eq~\ref{e21}). Here, the entropy is the dominant part, and it favors larger $z$ as $\phi$ increases. This can be understood because the $p$ polymers are more likely to adsorb an ion and become charged if the system is denser and the counterions are close to the polymers.

In the ACR model, $z$ is regulated by $\phi$ in a more pronounced way than in the HCR model, as seen by comparing Figure~\ref{fig2}a,b. In addition, Figure~\ref{fig2} shows that when  $\alpha$ decreases, $z$ will increase, as expected from the CR process. This tendency applies also when $\eta$ is decreased (not shown in the figure).

Special attention should be given to the different behavior in the dilute limit ($\phi\to 0$) of the two models. In the HCR (when $\eta=0$) the charge density $z$ approaches a constant value,
\be
\label{e212}
\begin{aligned}
\lim_{\phi\to 0} z/z_0 = 1/(1+{\rm e}^\alpha).
\end{aligned}
\ee
As no free ions exist in the HCR model, the limit for $\phi \to 0$ is different for different $\alpha$ values. For the ACR model, on the other hand, the dilute $\phi$  limit leads to $z\to 0$. This is due to the large entropy that the counterions gain as they remain dispersed in the solution.

\begin{figure}[h]
\centering
{\includegraphics[width=0.45\textwidth,draft=false]{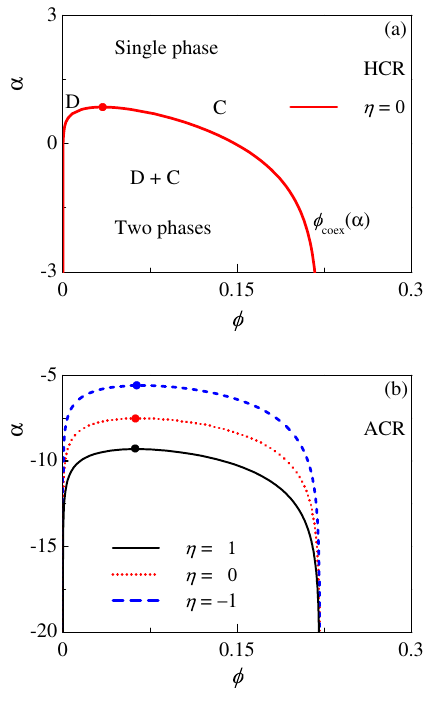}}
\caption{
\textsf{($\phi, \alpha$) phase diagram shown for different values of $\eta$. Below the coexistence curve $\phi_{\rm coex}(\alpha)$, the condensed (C) and dilute phases (D) coexist. (a) HCR for $\eta=0$ and (b) the ACR model for $\eta=1,0$, and $-1$. The critical point on each coexistence curve $\phi_{\rm coex}(\alpha)$ is denoted by a dot. The other parameters are $\gamma=0.2$, $\chi=0.5$, and $\lambda=26.68$.}}
\label{fig3}
\end{figure}
%
\subsection{Effects of CR on the Phase Separation}
We first present the effect of $\alpha$ on the phase diagrams, recalling that the $\alpha$  parameter quantifies the free-energy change of single-ion adsorption. Figure~\ref{fig3} shows the ($\phi, \alpha$) phase diagram for both the HCR and ACR models. A typical phase diagram is shown in Figure~\ref{fig3}a for the HCR model, where the phase separates into two coexisting polymer phases: a dilute phase (D) and a condensed phase (C). The phase separation occurs below the coexistence (binodal) curve, $\phi_{\rm coex}(\alpha)$, which terminates at an upper critical point, $(\alpha_c, \phi_c)$, marked by a dot in Figure~\ref{fig3}.

\begin{figure}[h]
\centering
{\includegraphics[width=0.45\textwidth,draft=false]{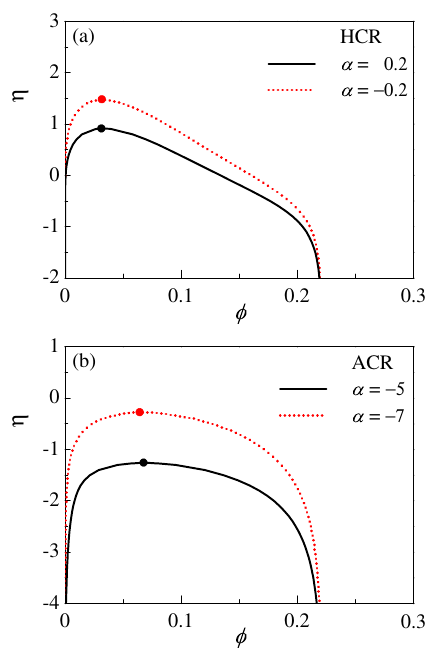}}
\caption{
\textsf{$(\phi,\eta)$ phase-diagram for (a)~the hopping CR model (HCR) for $\alpha=0.2$ and $-0.2$, and (b)~the asymmetric CR model (ACR) for $\alpha=-5$ and $-7$. The other parameters are $\gamma=0.2$, $\chi=0.5$, and $\lambda=26.68$.
}}
\label{fig4}
\end{figure}

In Figure~\ref{fig3}a,b, we see that decreasing $\alpha$ toward more negative values enlarges the gap between the volume fractions of the two coexisting phases, $\phi_1(\alpha)$ and $\phi_2(\alpha)$. We conclude that although the two models present very different CR mechanisms, they both show that decreasing $\alpha$ enhances the phase separation. In addition, the effect of $\eta$ on the phase diagrams for the two models exhibits behavior similar to that of $\alpha$. As shown in Figure~\ref{fig4}, decreasing $\eta$ will enlarge the polymer concentration asymmetry of the two coexisting phases.

\begin{figure}
\centering
{\includegraphics[width=0.45\textwidth,draft=false]{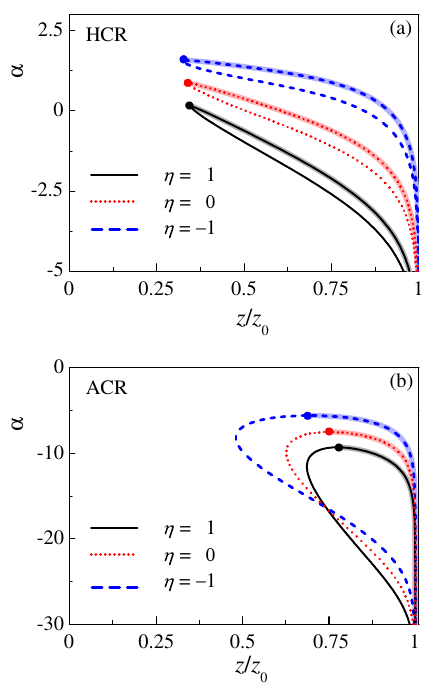}}
\caption{
\textsf{$(z/z_0,\alpha)$ phase diagram shown for different $\eta$ values for (a)~the HCR model and (b)~the ACR model. In both (a) and (b),  $\eta=1, 0$, and $-1$. The other parameters are $\gamma=0.2$, $z_0=\gamma N=40$, $\chi=0.5$, and $\lambda=26.68$. The thick lines with a shadow correspond to the coexisting condensed (C) phase.}}
\label{fig5}
\end{figure}

Next we investigate the effect of the two CR parameters on the polymer ionization state in the two phases. In Figure~\ref{fig5} we show the phase diagram in the ($z/z_0, \alpha$) plane by imposing the relation $z(\phi)$ presented in Figure~\ref{fig2} on the ($\phi, \alpha$) phase diagram. A related phase diagram in the ($z/z_0, \eta$) plane is shown in Figure~\ref{fig6}.
In both the HCR and ACR models, the polymer chains are more charged in the C phase than in the D phase as can be seen in Figure~\ref{fig5}, consistent with the $z(\phi)$ relations presented before.

We note that this result is {\it opposite} to the conclusions in ref~\cite{Muthukumar2010}, where the condensed (C) phase was less charged
than the dilute (D) phase. The difference stems from the details of the ionization process of the polymers. In their study~\cite{Muthukumar2010}, the polymers became
charged by releasing ions into the solution. Hence, this guarantees that at extreme dilution the polymer charge would be maximal.
In our ACR model, the charging mechanism of the $p$ polymers is opposite and consists of adsorbing ions from the solution. This leads to a higher charge at
large densities, 
in agreement with the general tenets of Le Chatelier's principle.
Finally, in our HCR model, no counter-ions are present and a comparison to ref~\cite{Muthukumar2010} is harder to make. The condensed phase in the HCR model is the more charged one because it is electrostatically favorable, as discussed in section III.A with regards to the $z(\phi)$ relation.

Figure~\ref{fig5} shows an important difference in the CR induced phase separation of the two models, HCR and ACR. For the HCR model (Figure~\ref{fig5}a), a smaller and more negative $\alpha$ increases the polymer charge in both the dilute (D) and condensed (C) phases, keeping the charge difference between the two phases relatively small. As a result, the two phases in the HCR model have distinct polymer densities but similar charges.
For the ACR model (Figure~\ref{fig5}b), on the other hand, the charge density of the polymers in the two phases can differ substantially, and nonmonotonic behavior is observed. As $\alpha$ decreases from $\alpha_c$, the polymer charge in the dilute (D) phase decreases first, and then increases, and the polymer charge in the condensed (C) phase monotonically increases. In other words, when the CR parameter $\alpha$ changes in a way that favors ion adsorption, it increases the charge asymmetry at first and then decreases it.

The novel effect of CR on the polymer charge presented in Figure~\ref{fig5} can be explained as a competition between direct and {\it secondary CR effects}. The direct effect, present also in the stable single phase, means that $z$ becomes larger as $\alpha$ decreases. This is clear from the CR mechanism and also is shown in Figure~\ref{fig2} for both models. On the other hand, decreasing $\alpha$ makes the dilute (D) phase more dilute, as seen in Figure~\ref{fig3}, and from the $z(\phi)$ relation in Figure~\ref{fig2}, the dilution causes $z$ to decrease. This change in $z$ is a secondary CR effect, as it involves the effect of CR first on $\phi$, and then on the charge. Note that for the HCR model the regulation of $z$ from the change in $\phi$ is minor. Therefore, the secondary CR effect is negligible and the charge density increases in both phases. For the ACR, $z$ is regulated by $\phi$ in a pronounced way, causing the two mentioned effects to be comparable, and results in the nonmonotonic behavior as observed.

Finally, Figure~\ref{fig5} shows the effect of the second CR parameter, $\eta$ that quantifies the interaction between different adsorption polymer sites, on the ($z/z_0, \alpha$) phase diagram. For the HCR model, decreasing $\eta$ increases the charge of both dilute and condensed (C) phases, as seen from looking at a fixed $\alpha$ value in Figure~\ref{fig5}, for different $\eta$ values.
However, for the ACR model, a change in $\eta$ causes a nonmonotonic behavior.
One can see that for a fixed $\alpha \approx \alpha_c$, decreasing $\eta$ causes the polymer charge in the dilute (D) phase to decrease and in the condensed (C) phase to increase. Hence, it increases the charge asymmetry between the two phases. As $\alpha$ becomes more negative, this effect becomes smaller until the trend reverses and the decrease in $\eta$ causes a higher polymer charge in the two phases and lowers the polymer charge asymmetry. This nonmonotonicity stems from secondary CR, as explained in the previous paragraph.

\section{Conclusions}

In summary, we study the effect of the charge regulation (CR) mechanism on the complex coacervation phase separation. Specifically, we considered two variants of the CR model: (i)~the hopping CR model (HCR) and (ii)~the asymmetric CR model (ACR). We introduce two CR parameters:  the association-dissociation energy  parameter of a single adsorption site $\alpha$, and the short-range nearest-neighbor interaction strength between the occupied sites along the polymer chain, $\eta$. The effects of the two CR parameters on the phase diagram have been studied in detail for the two models. When either $\alpha$ or $\eta$ is decreased, the tendency  to phase separate increases. This trend can be tested in experiments where the acid dissociation constant is varied, either by using different types of polyelectrolytes or by controlling chemically-grafted ionic groups on the polyelectrolyte chains.

An important conclusion that has yet to be verified in experiments is the following. The polymer charge in the two phases is regulated directly by the chemical reactions that determine the charge in the single phase, as well as indirectly because the CR changes the volume fraction of the phases, which in turn regulates the polymer charge even further. The two competing CR effects can cause a nonmonotonic behavior of the charge asymmetry between the two phases as function of the CR parameters.

We hope that the charge regulation mechanism as explored in this work will provide insight into the understanding of the complex coacervation in experiments on biological and synthetic materials.

\textit{\textbf{Acknowledgements}}:
This work was supported by the National Natural Science Foundation of China (NSFC) -- the Israel Science Foundation (ISF) joint program under grant no. 3396/19, and by the ISF
under grant no. 213/19.  Y.A. is thankful for the support from the Clore Scholars
Programme of the Clore Israel Foundation.
R.P. acknowledges the support of the University of Chinese Academy of Sciences and the NSFC under Grant No.12034019.

\section*{Appendix}

\subsection{$(z/z_0,\eta)$ phase diagram}\label{AppendixA}

\begin{figure}[h]
\centering
{\includegraphics[width=0.45\textwidth,draft=false]{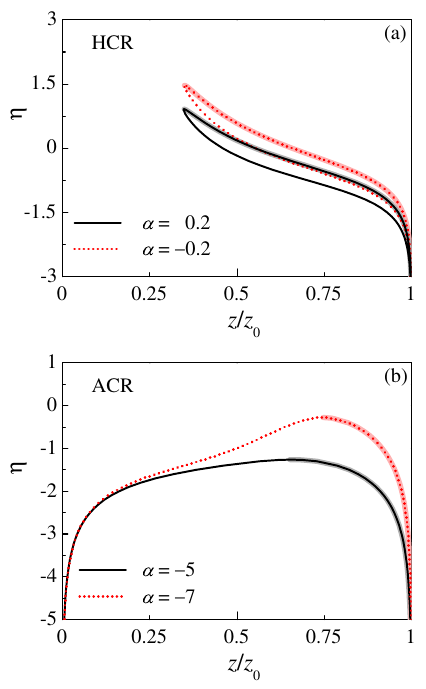}}
\caption{
\textsf{$(z/z_0,\eta)$ phase diagram. (a) $\alpha=0.2$ and $-0.2$ for the HCR model. (b) $\alpha=-5$ and $-7$ for the ACR model.
The other parameters are $\gamma=0.2$, $z_0=\gamma N=40$, $\chi=0.5$, and $\lambda=26.68$. The thick lines with a shadow correspond to the condensed phase (C).
}}
\label{fig6}
\end{figure}

Figure~\ref{fig6} presents the effect of the $\eta$ parameter on the polymer ionization state in the two phases. For the HCR model shown in Figure~\ref{fig6}a, the polymer charge in both the dilute (D) and condensed (C) phases increases as $\eta$ decreases. For the ACR model, in the range of $\alpha\approx \alpha_c$, the decrease in $\eta$ causes the polymer charge to increase in the condensed phase (C) and decreases in the dilute phase (D) as shown in Figure~\ref{fig6}b. For negative, large-enough $\alpha$, the polymer charge in both phases increases as $\eta$ increases. This tendency is not presented in Figure~\ref{fig6}b but is shown in Figure~\ref{fig5}b. This non-monotonicity originates from the secondary CR effect, as explained in section~III.B.

\newpage

\clearpage

\newpage
\vskip 0.5truecm
\centerline{For TOC Graphics}
\centerline{\bf Phase Separation of Polyelectrolytes: }
\centerline{\bf the Effect of Charge Regulation}
\centerline{\it Bin Zheng, Yael Avni, David Andelman, and Rudolf Podgornik}

\begin{figure}[h]
{\includegraphics[width=0.4\textwidth,draft=false]{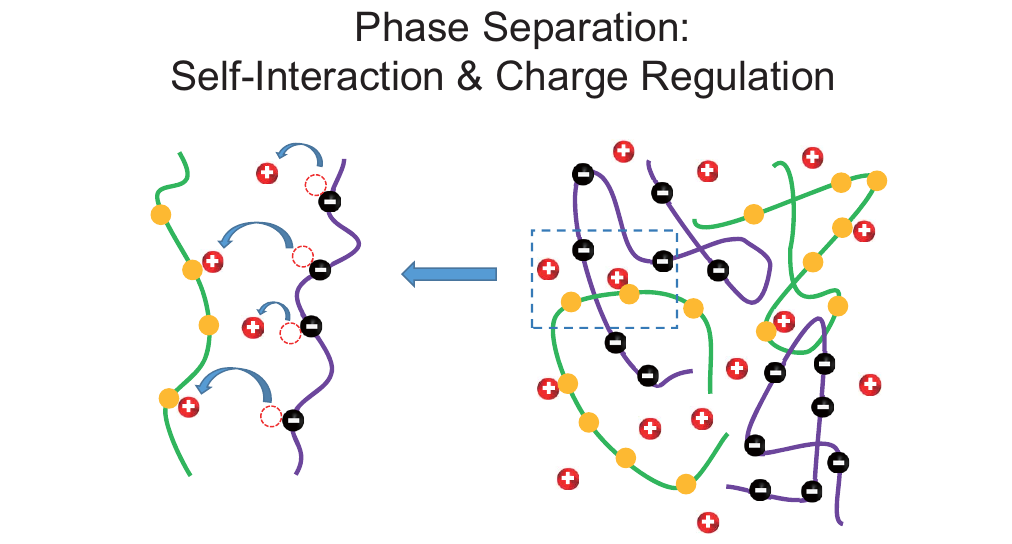}}
\label{TOC Graphic}
\end{figure}

\end{document}